\documentclass[sigconf]{acmart}

\AtBeginDocument{%
  }

\copyrightyear{2024}
\acmYear{2024}
\setcopyright{rightsretained}
\acmConference[CIKM '24]{Proceedings of the 33rd ACM International Conference on Information and Knowledge Management}{October 21--25, 2024}{Boise, ID, USA}
\acmBooktitle{Proceedings of the 33rd ACM International Conference on Information and Knowledge Management (CIKM '24), October 21--25, 2024, Boise, ID, USA}
\acmDOI{10.1145/3627673.3680035}
\acmISBN{979-8-4007-0436-9/24/10}

\begin{document}

\title{Ads Supply Personalization via Doubly Robust Learning}


\author{Wei Shi}
\affiliation{%
  \institution{Meta Platforms, Inc.}
  \city{Sunnyvale}
  \state{CA}
  \country{USA}
}
\email{weishi0079@meta.com}

\author{Chen Fu}
\affiliation{%
  \institution{Meta Platforms, Inc.}
  \city{Sunnyvale}
  \state{CA}
  \country{USA}
}
\email{chenfu@meta.com}

\author{Qi Xu}
\affiliation{%
  \institution{Meta Platforms, Inc.}
  \city{Sunnyvale}
  \state{CA}
  \country{USA}
}
\email{xuqi@meta.com}

\author{Sanjian Chen}
\affiliation{%
  \institution{Meta Platforms, Inc.}
  \city{Menlo Park}
  \state{CA}
  \country{USA}
}
\email{sjchen@meta.com}

\author{Jizhe Zhang}
\affiliation{%
  \institution{Meta Platforms, Inc.}
  \city{Menlo Park}
  \state{CA}
  \country{USA}
  \country{}
}
\email{jizhezhang@meta.com}

\author{Qinqin Zhu}
\affiliation{%
  \institution{Meta Platforms, Inc.}
  \city{Menlo Park}
  \state{CA}
  \country{USA}
}
\email{catherinezhu@meta.com}

\author{Zhigang Hua}
\affiliation{%
  \institution{Meta Platforms, Inc.}
  \city{Sunnyvale}
  \state{CA}
  \country{USA}
}
\email{zhua@meta.com}

\author{Shuang Yang}
\affiliation{%
  \institution{Meta Platforms, Inc.}
  \city{Sunnyvale}
  \state{CA}
  \country{USA}
}
\email{shuangyang@meta.com}








\renewcommand{\shortauthors}{Wei Shi et al.}

\begin{abstract}
Ads supply personalization aims to balance the revenue and user engagement, two long-term objectives in social media ads, by tailoring the ad quantity and density. In the industry-scale system, the challenge for ads supply lies in modeling the counterfactual effects of a conservative supply treatment (e.g., a small density change) over an extended duration. In this paper, we present a streamlined framework for personalized ad supply. This framework optimally utilizes information from data collection policies through the doubly robust learning. Consequently, it significantly improves the accuracy of long-term treatment effect estimates. Additionally, its low-complexity design not only results in computational cost savings compared to existing methods, but also makes it scalable for billion-scale applications. Through both offline experiments and online production tests, the framework consistently demonstrated significant improvements in top-line 
business metrics over months. The framework has been fully deployed to live traffic in one of the world's largest social media platforms.
\end{abstract}







\maketitle

\section{Introduction}

\label{sec:intro}

\begin{figure}[!htb]
    \centering
    \includegraphics[width=0.5\textwidth]{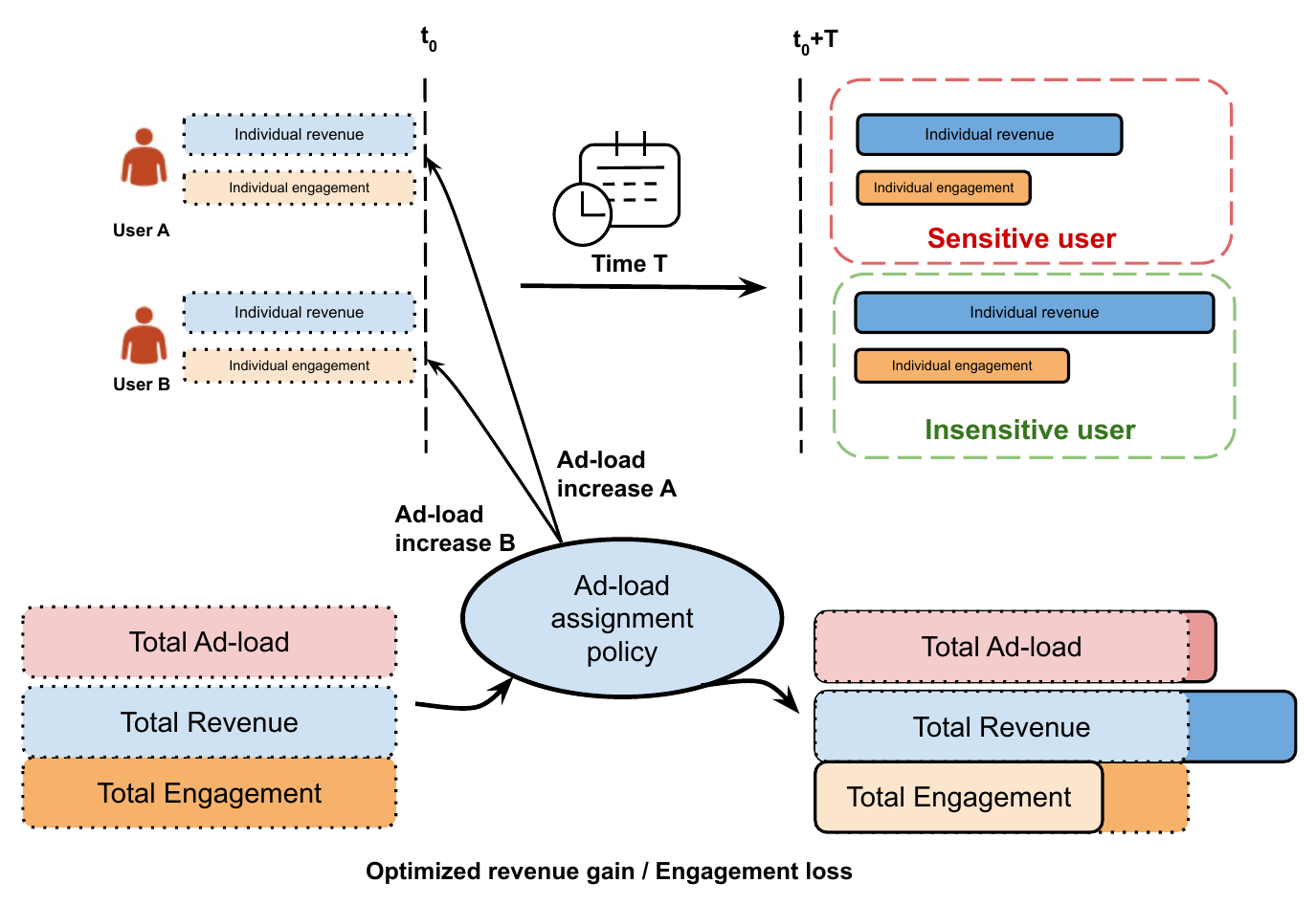}
    \caption{Systematic overview of ads-supply personalization.}
    \label{fig:ad_supply_system}
\end{figure}
Social advertising platforms generate revenue by ads and user engagement by organic content. Balancing user engagement and revenue generation is important. Typically, boosting ad delivery enhances short-term revenue but may lead to a decline in user engagement, impacting long-term revenue prospects adversely. Addressing the optimal trade-offs between these conflicting objectives represents a significant challenge in the realm of online advertising.
\\
Personalization is one of the important strategies that has proven to be effective in boosting revenue and engagement separately, typically by customizing content selection and ranking. Yet, the separate personalization strategy does not take into account the often-conflicting nature and intricate interplay between the two objectives (i.e, revenue and engagement ). 
To attain optimal trade-offs, there is a crucial need for a joint personalization strategy that delves deeper into customizing the blending algorithm itself. Along this line, existing efforts have largely focused on ads allocation, specifically by customizing the positions of ads and organic content, for example, via constrained optimization \cite{yan2020ads} and Markov decision process (MDP) \cite{liao2022cross, liao2022deep, wang2022hybrid, zhao2021dear, xue2022resact}. However, there still exists a significant gap in research concerning the optimization of ad quantity and density, which directly influence top-line revenue and engagement metrics, making further exploration imperative from an industry perspective.
\\
In this paper, we examined a personalized strategy designed to control the level of ad-load to balance revenue and user engagement, termed as \textit{ads supply personalization}. 
The effectiveness of personalization arises from tailoring ad load to individual users according to their perceived value of the ads content, rather than applying an one-size-fits-all configuration. Such heterogeneous ads perceptions are quantified by users responses (e.g, revenue gain and engagement loss) from an ad-load treatment (e.g, an increased load level). We define users with high revenue increase and low engagement loss as insensitive users. A selective ad-load increase with insensitive users well identified can significantly improve the cumulative revenue gain-to-engagement loss efficiency over a uniform ad-load increase as shown in Figure \ref{fig:ad_supply_system}. 
\\
In the billion-scale industrial applications, the key challenge of ads supply personalization is to build a model for users' long-term causal effects from an ad-load change. Long-term ads-supply effects are difficult to estimate because of the small-magnitude treatment and complicated observations in a long time window. Firstly, ad-load changes, the treatments, are usually small to be conservative since they directly impact top-line business metrics. As a result, the impact (i.e, to revenue and engagement metrics) is so tiny that distinguishing it from background noise becomes a real challenge. Secondly, to measure long-term impact, user responses are often monitored over an extended period (e.g., months) to ensure consistency, which introduces additional challenges and complexities. For example, observations on revenue and engagement can easily be overshadowed by various unrelated factors such as seasonality and ongoing changes made into  production (e.g., the launch of new ranking models). Furthermore, ads supply personalization should be low-latency and less computationally intensive to ensure minimal additional infrastructure costs on top of the established ads and organic content delivery systems. 
\\
Existing works either directly reuse estimated utilities \cite{yan2020ads, carrion2021blending} like ads-CTR from ranking models or build the model on users' immediate and short-term signals \cite{liao2022cross, liu2023explicit, wu2023dnet}. Modeling long-term user effect, which is crucial for ads supply personalization, remains largely unexplored. 
These models for short-term causal effect modeling are also difficult to scale up as they are already computationally demanding due to deployments of dedicated deep neural networks (DNN) \cite{liu2023explicit, wu2023dnet}. 
\\
In this paper, we present a streamlined framework based on doubly robust learner (DRL), which provides a lightweight solution to model the long-term causal effects in large-scale ads supply personalization. The improved design on utilization of information during data collection and modeling stages leads to substantial enhancements in performance and reductions in model complexity. Our framework can be easily integrated with separate ad and organic delivery systems, resulting in minimal increases in computational complexity. This framework offers a comprehensive optimization across stages of data collection, modeling, and deployment, grounded in a theory from causal inference.
The contributions of our work are summarized as follows:
\begin{itemize}
    \item \textbf{A lightweight ads-supply personalization for long-term metrics.} In this paper, we propose a new framework based on DRL to model the long-term causal effects from ad-load changes. The framework fully leverages the information of random-controlled trial (RCT) data to significantly improve the quality of treatment effect estimates at a reduced model complexity. We present detailed analysis under DRL formulation and design a lightweight implementation for the large-scale application correspondingly.
    
    \item \textbf{Detailed industrial and practical experience.} We conducted extensive empirical studies on the public and real-world benchmarks. The proposed system and the reported improvements have been fully deployed on with live traffic in one of the world's largest social media platforms.
\end{itemize}

\section{Preliminaries and Related Works}
\subsection{Uplift Modeling}
Uplift models have been extensively used in heterogeneous treatment effect (HTE) or conditional average treatment effect (CATE), which is widely applicable in medical sciences, psychology, sociology, and economics. Following the Neyman-Rubin potential outcome framework \cite{rubin2005causal}, the observed responses are the outcome and their incremental changes of outcome are the treatment effect. There are three major categories of uplift models. (1) Causal Trees. Tree-based models divide the user population using different splitting criteria, such as distribution divergences \cite{radcliffe2011real} and expected responses \cite{saito2020cost, zhao2017uplift}. The extension of causal trees is causal forest \cite{athey2016recursive}. (2) Meta-learner. Meta-learner \cite{kunzel2019metalearners} tackles the counterfactual problem by predicting the potential outcomes. (3) Deep Neural Network (DNN) causal models. Strong predictive performance and representation power of DNNs motivate researchers to work on deep causal models \cite{shalit2017estimating, johansson2016learning, louizos2017causal, alaa2017deep, schwab2018perfect, yoon2018ganite, farrell2021deep}. The consistency of treatment estimates from meta-learner and deep causal models highly rely on the consistent outcome models. We will show that it is not satisfactory to build a complex outcome model for the long-term treatment effects in our ads-supply personalization. Instead, we can build a doubly robust learner on RCT data.

\subsection{Doubly Robust Learning}
Doubly Robust Learning \cite{foster2023orthogonal, kennedy2023towards} is a method for estimating CATE even if either the propensity score or outcome modeling cannot be satisfactorily modeled by parametric functions. Doubly robust learning works under the conditions of unconfoundedness \cite{rosenbaum1983central}, where all potential confounders are observed, and categorical treatments. The methodology has been developed primarily for average treatment effect (ATE) \cite{glynn2010introduction, van2006targeted, yang2023multiply, tan2022doubly}. Recently doubly robust approach is combined with machine learning to estimate CATE like DRL in \cite{kennedy2023towards}. DRL illustrates a application scenario that we can leverage a known propensity score to achieve an unbiased CATE estimates if the individual outcomes are difficult to estimate on their own but CATE is more structured. Therefore, we adopt the DRL methodology to avoid the resource intensive outcome model in ads-supply modeling and still obtain satisfactory CATE estimates. 

\subsection{Ads Allocation}
\label{subsec:auuc}
A related problem is the ads allocation optimization. The ads allocation not only determines whether to insert a ad in the organic content flow, but also determines the optimal position. A blending layer for ads and organic contents can be added and formulated as a constrained optimization problem \cite{yan2020ads}, or multi-objective optimization \cite{carrion2021blending}. Their focus is the optimization on positions based on existing utility signals like ad-CTR in \cite{carrion2021blending}. Since the optimization is conducted at the session-level, the work of \cite{yan2020ads} also sets the guardrails including top slot and min gap to avoid disastrous experience in the user level. Sophisticated DNN modules been extensively explored in the ads allocation\cite{liao2022cross, liao2022deep, wang2022hybrid, zhao2021dear, xue2022resact}.
\\
Our ads-supply personalization distinguish the above problem formulation in two folds: (1) we explicitly control the user-level ads quantity. The existing ads allocation either leave ads quantity as an uncontrolled confounder or use a pre-defined quantity threshold \cite{carrion2021blending}. (2) We model the long-term metrics as utilities for optimization while above works either leverage existing utilities or immediate user signals \cite{liao2022cross}. 

\section{Problem Formulation}
\subsection{Constrained Optimization}
One business goal is to optimize for the revenue gain at a fixed engagement loss budget. The ad-load increase is discretized into a binary treatment so that the constrained optimization problem falls into the category of 0-1 knapsack problem:

\begin{equation}
     max\sum\limits_{i=1}^{n} \tau^r(x_i)z_i \quad
     s.t. \sum\limits_{i=1}^{n}
     \tau^e(x_i)z_i \leq B, z_i \in \{0, 1\}
     \label{eq:knapsack}
\end{equation}
\\
where $z_i \in \{0, 1\}$ is the treatment variable for $i^{th}$ user. In this context, $z_i=1$ signifies that an ad-load increase is assigned to the user, while $z_i=0$ indicates the absence of such an increase. The terms $\tau^r$ and $\tau^e$ represent the treatment effects for revenue gain and engagement loss, respectively. As users have different tolerance to ad-load increase, quantified by their treatment effect ratio $\tau^r/\tau^e$, a personalized ad-load assignment will benefit the cumulative revenue gain. If $\tau^r$ and $\tau^e$ are known, the optimal solution turns out to be the greedy algorithm of selecting the largest $\tau^r/\tau^e$ since $\tau^e$ is much smaller than the total budget \cite{dantzig1957discrete}. The key challenge lies in building a model for estimating the treatment effects of $\tau^r$ and $\tau^e$. Therefore, we adopt the uplift model, a category of models in causal inference with an emphasis on modeling heterogeneous treatment effects.
\subsection{Heterogeneous Treatment Effect Modeling}
Our primary modeling objective, $\tau^r$ and $\tau^e$, are long lasting effects from a conservative ad-load change of small magnitudes. We formalize our modeling problem as a heterogeneous treatment effects (HTE), or conditional average treatment effect (CATE) estimate problem. We follow the potential outcome framework\cite{rubin1974estimating} in the causal inference. We have $n$ independent and identically distributed samples $(X_i, Y_i, W_i), i=1,...,n$, where $X \in \chi$ are per-sample features, $Y_i \in \mathbb{R}$ is the observed outcome, and $W_i \in {0, 1}$ is the treatment assignment indicator. The potential outcomes ${Y_i(0), Y_i(1)}$ are the outcome we would have observed given the treatment $W_i=0$ or $1$ respectively. Our objective of interest is to estimate CATE:

\begin{equation}
\tau(x) = \mathbb{E}[Y(1) - Y(0)|X=x] \label{eq:CATE}
\end{equation}
\\
In our problem, we set up two separate CATE models, one for revenue and the other for engagement treatment effects. The specific meaning for each variable mentioned above is specified below.
\\
\textbf{Outcome $\textbf{Y}$} is the long-term user revenue or engagement metrics. These metrics are gathered over an extended period, during which numerous perturbations may occur. Consequently, the association between features and outcomes are challenging to model as they are dominated by the large noises.
\\
\textbf{Treatment $\textbf{W}$} is the user-level ad-load change that influences both revenue and engagement metrics. The ad-load change is discretized into a binary variable $W \in \{0, 1\}$. The ad-load changes has a clear semantic meaning like top slot and min gap in \cite{yan2020ads} and a small magnitude for conservative business decisions.
\\
\textbf{Feature $\textbf{X}$} is the pre-treatment user features like user characteristics and activity loggings. The feature values should not be impacted by the treatment after model launched online. Otherwise, a potential positive feedback loop could lead to a disaster in the ad-load assignment. 
\\
\textbf{CATE $\tau$} is the long-term and incremental changes in revenue and engagement for a certain user cohort. The individual revenue-to-engagement ratio, $\tau^r / \tau^e$, is defined as user's sensitivity to ads. The larger the ratio is, the more insensitive user is.
\subsection{Challenge for Long-term Effect Estimate}
Popular uplift models like meta-learners in \cite{kunzel2019metalearners} rely on outcome modeling for individual-level CATE estimates. However, our outcomes are long-term revenue and engagement metrics, whose association with features are very difficult to capture. This issue presents a dilemma: we must either accept an unsatisfactory CATE estimate or bear with the prohibitive complexity of the model. To address this challenge, we first consider the full utilization of information in both the data collection and modeling stages. Subsequently, we devise a strategy to create a lightweight CATE model that is compatible with the constrained online serving environment.
\\
\textbf{Nuisance modeling} Because we do not have the true labels for CATE due to the counterfactual problem, outcomes are first predicted. Outcome predictions are only used for estimating CATE, hence defined as the nuisance. Therefore, a straightforward CATE estimator is:
\begin{equation}
    \hat{\tau}(X_i) = \hat{Y}(X_i, W=1) - \hat{Y}(X_i, W=0) \label{eq: naive_estimator}
\end{equation}
where $\hat{Y}(X_i, W)$ is the prediction from the outcome model. However, such method is not satisfactory when the outcomes are hard to estimate on their own. Furthermore, it is particularly inefficient to build a complex outcome model to estimate a more structured CATE function.
\\
\textbf{Data collection} Although many methods are designed to learn from observational data \cite{kunzel2019metalearners, shi2019adapting, louizos2017causal}, using RCT data as training data can make the modeling easier. RCTs are the gold standard for ascertaining the efficacy and safety of a treatment so that they are applied in large-scale applications \cite{liu2023explicit, wu2023dnet}. However, the existing works focus more on building capable outcome modeling while not fully leveraging the information from RCTs.
\\
As a result, the strategy for long-term CATE estimates should aim to (1) maximize the extraction of information from RCT data collection to enhance performance, and (2) separate a lightweight CATE model from nuisance models for online serving. The detailed methodology is outlined in the subsequent section.

\section{Methodology}
\subsection{Doubly Robust Learner with RCT}
We collect training data from a RCT. Users are first divided into treatment and control groups. The labels collected in the RCT will be modified by a long-term (4 months+) projection. Usually, unconfoundedness is required to identify the treatment effect. 

\begin{equation}
    \{Y_i(0), Y_i(1)\}\bot W_i|X_i \label{eq:unconfounderness}
\end{equation}
\\
RCT data actually poses a stronger assumption than unconfoundedness:

\begin{equation}
     \{Y_i(0), Y_i(1)\}\bot W_i \label{eq:independence}
\end{equation}
\\
Therefore, RCT provides us with additional information, true propensity score, besides the observed outcomes. Here propensity score refers to the probability of receiving treatment. We now denote the propensity score as $e(x)=p$ where $p$ is the overall treatment percentage in the RCT.
\\
As mentioned, the direct estimator in \eqref{eq: naive_estimator} is highly dependent on the consistency of outcome model $\hat{Y}(x)$. It is not consistent if the outcome model is misspecified. We can address such problem by fully harnessing the true propensity score.   The doubly robust (DR) CATE estimator is constructed by:

\begin{align}
    \hat{\tau}(X_i) = \hat{Y}^{DR}(X_i, W=1) - \hat{Y}^{DR}(X_i, W=0)
    \label{eq: dr_estimator}
\end{align}

where:

\begin{align}
     \hat{Y}^{DR}(X_i, W=t) = \hat{Y}(X_i, W=t) + \frac{Y_i - \hat{Y}(X_i, W=t)}{e_t(X_i)} \cdot 1\{W_i=t\}
\end{align}
\\
With a known $e(x)=p$, DR estimator guarantees the consistency even if the outcome model is misspecified. In essence, the DR estimator maximizes the use of information available in RCT data to lessen the demands on the outcome model. Building on the DR estimator, DRL provides a distinct CATE model separate from the outcome model. In addition to eliminating the outcome bias inherent in the DR estimator, the standalone CATE model in DRL is not burdened by the complexity of the outcome model during online serving.
\begin{figure}[!htb]
    \centering
    \includegraphics[width=0.35\textwidth]{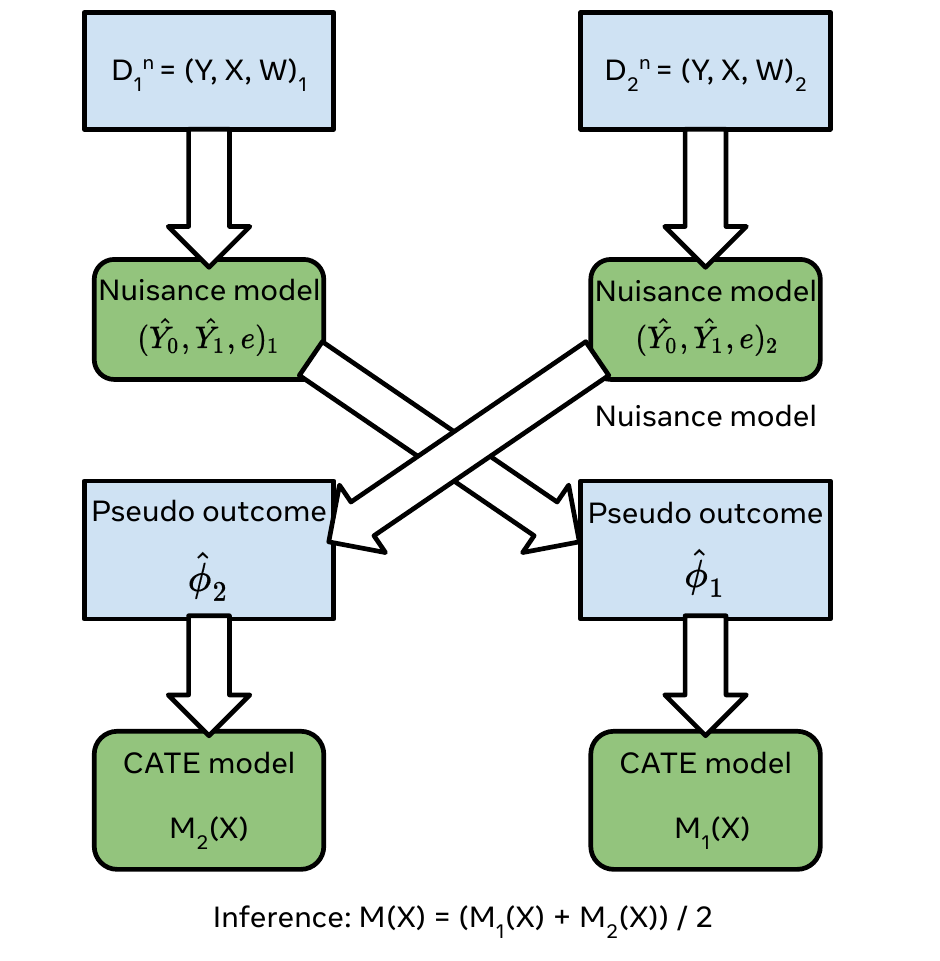}
    \caption{The flow of doubly robust learner.}
    \label{fig:drl}
\end{figure}
\\
Figure \ref{fig:drl} shows the DRL flow consisting of two predictive tasks. First, we split the training data into two subsets $D_1^n$ and $D_2^n$. Then potential outcome predictions (nuisance) $\hat{Y}_0, \hat{Y}_1$ are obtained by regression on the observed outcome $Y$. With a known propensity score $e(X)=p$ from RCT, we construct the pseudo-outcome by:

\begin{equation}
     \hat{\phi}(X) = \frac{W - p} {p \{1 - p\}}  \cdot \{Y - \hat{Y}(X, W)\} + \hat{Y}(X, W=1) - \hat{Y}(X, W=0)\
\end{equation}
\\
The first term on the right-hand side represents the integration of treatment information in the RCT data. This ensures the consistency of the CATE estimator, even in the presence of biased outcome predictions. At last, we construct the second predictive model by regressing $\hat{\phi}$ on $X$, resulting in a CATE model $M(X)$. 
\\
\subsection{A Lightweight Implementation}
\textbf{Nuisance and CATE regressor} Since the bias of outcome predictions is not carried over to CATE estimator in the combination of DRL and RCT, the requirement on nuisance model is largely reduced. We implement the outcome models with random forest, which significantly saves the computation power compared to DNN implementations in \cite{liu2023explicit, wu2023dnet}. We also implement the second-stage CATE regressor by random forest to satisfy the stability condition as outlined in \cite{kennedy2023towards}. 
\\
\textbf{High-dimensional feature consumption} In order to augment the predictive capabilities of the CATE model based on random forest, we have also devised methods to process high-dimensional features. One such significant feature is the user embedding, which is generated from other large pre-trained models. We utilized unsupervised learning to segregate users into distinct clusters based on their embeddings. The cluster ID is then used as a categorical feature in the random forest model, leading to an enhancement in performance.
\\
\textbf{Feature selection} We need to choose features that directly contribute to CATE estimates from numerous candidates. For each feature, users are divided into different cohorts according to the feature value. Consequently, an AUUC value can be derived from ground-truth CATE from each cohorts. Such single-feature AUUC is used to filter out most of candidates.
\\
\textbf{Lagrangian selecting criteria}
Although the ratio form, $\tau^r/\tau^e$, is ideally optimal in \eqref{eq:knapsack} as a greedy selection criteria, it is not the case in the real-world. The ratio form amplifies the prediction noises and loses some information. Instead, we adopted the Lagrangian selecting criteria \cite{du2019improve}, which is a linear form. Furthermore, it is also compatible with DR estimator.
\begin{equation}
     S_i = \hat{\tau}^r - \lambda \hat{\tau}^e
\end{equation}
where we treat $\lambda$ as another hyper-parameter of the model and use parameter search to determine its value. In our actual implementation, we fuse separate revenue and engagement CATE regressors into a unified stage to generate the score $S_i$. Furthermore, we apply clustering to raw scores to further reduce variance. An alternative way is to build a direct ranking model (DRM) in \cite{du2019improve}. However, DRM demands non-trivial computing resources at the large scale, which is a less ideal candidate for the lightweight solution to our problem.
\section{EMPIRICAL EVALUATION}
In this section, we conduct experiments to evaluate our framework for large-scale industrial applications and present analysis for systematic trade-offs. Our experiments have the following objectives:
\begin{enumerate}
  \item Evaluate the end-to-end improvement in CATE estimate accuracy in public datasets and business metrics in product datasets, brought about by our framework.
  \item Demonstrate the rationale behind our system design through component-level experiments and showcase the adaptability of our framework with larger datasets.
  \item Validate the long-term impacts of our framework on the products, both by a one-month pre-launch A/B test and N-month post-launch hold-out test (N $\geq$ 4).
\end{enumerate}
\subsection{RCT Datasets}
We evaluate the methods in both public and product datasets. The selected benchmarks meet the following criteria: (1) They are both large-scale advertisement datasets, and (2) They have been collected through RCTs, where all covariates are solely related to the outcome. The RCT setting guarantees that the propensity score can be known trivially. The public benchmark is CRITEO-UPLIFT \cite{diemert2021large}.
Our product dataset was collected through a RCT, recording users' long-term revenue and engagement metric with the pre-treatment feature values. The dataset contains tens of million user loggings with around twenty features. The ad-load change is a binary operation. We randomly split both two datasets into training data (80\%) and test data (20\%). 
\\
In these two RCT-based datasets, propensity score is the ratio of treatment group size over the entire population.


\subsection{Baseline Models}
We choose five popular methods that have been applied to the large-scale applications. They all rely on the outcome modeling to attain the treatment effect estimates. Here S/T/X-learner are based on random forest regressors. CEVAE and Dragonnet are DNN-based models. They represent different complexities of outcome models. Baseline methods are implemented by the CausalML package\cite{chen2020causalml}.
\\
\textbf{S-learner}
S-learner \cite{kunzel2019metalearners} is a single model to predict the outcome of individuals with the treatment indicator as a feature. 
\\
\textbf{T-learner}
T-learner \cite{kunzel2019metalearners} uses separate outcome models for treatment and control groups. It circumvents the bias towards zero in S-learner.
\\
\textbf{X-learner}
X-learner \cite{kunzel2019metalearners} is a two-stage approach. The first stage is the outcome model like T-learner. The second stage regresses the pseudo-outcome derived from data labels and first-stage predictions.
\\
\textbf{CEVAE}
CEVAE \cite{louizos2017causal} uses Variational Autoencoders (VAE) to simultaneously estimate the unknown latent space summarizing the confounders and the causal effect. The network structure focuses on the confounder modeling, which is an important aspect in causal representations for observational studies.
\\
\textbf{Dragonnet}
Dragonnet \cite{shi2019adapting} exploits the sufficiency of the propensity score for estimation adjustment by training the network for both propensity score and outcome predictions.

\subsection{Evaluation Metrics}
\begin{figure}[!htb]
    \centering
    \includegraphics[width=0.3\textwidth]{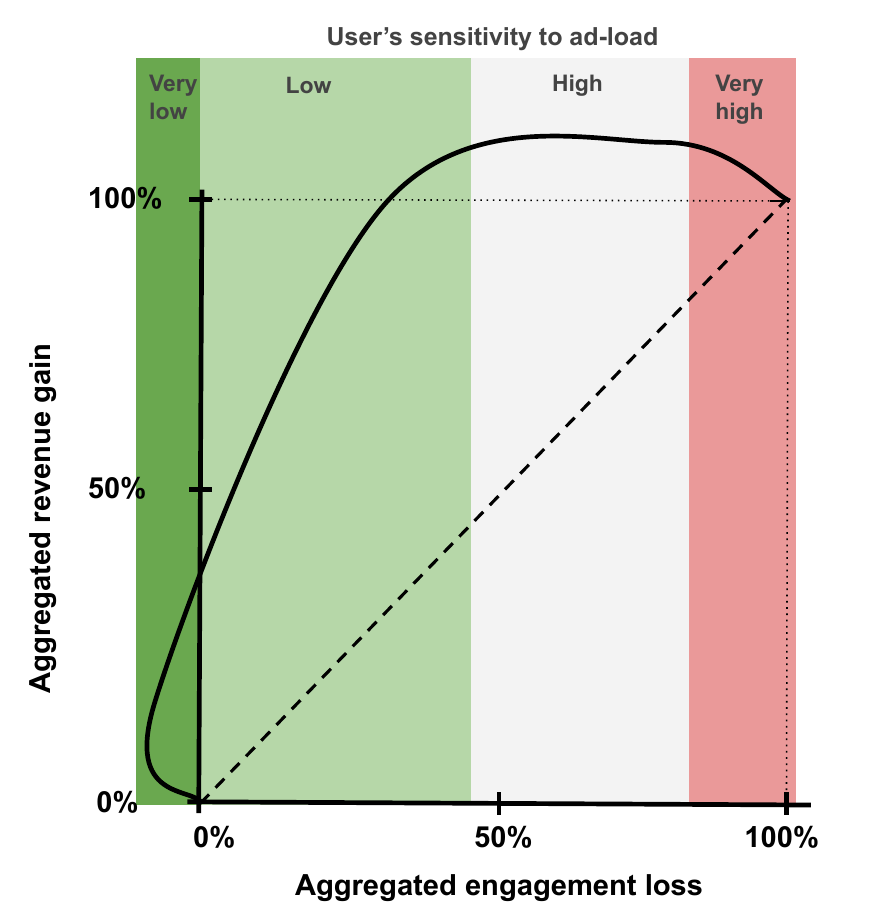}
    \caption{AUCC Definition in Ads Supply Personalization.}
    \label{fig:aucc}
\end{figure}

We use AUUC to measure the performance of a CATE model and AUCC to measure the business gain by a combination of revenue and engagement CATE models. 
\begin{itemize}
  \item[$\circ$] AUUC (Area under Uplift Curve) A common metric to evaluate heterogeneous effect predictions. This is used for CRITEO-UPLIFT as it has only one objective.
  \item[$\circ$] AUCC (Area under Cost Curve) \cite{du2019improve} An AUUC variant to evaluate the Return on Investment (ROI) predictions. The cost curve plots the aggregated incremental value as the Y-axis and the aggregated cost as the X-axis. In our problem, revenue gain is the value and engagement loss is the cost. AUCC is used for our product data as it is a dual-objective optimization. An cost uplift curve example is shown in Figure \ref{fig:aucc}. The more accurately we can identify insensitive users, the greater the efficiency we can achieve in terms of cumulative revenue gain relative to engagement loss.
\end{itemize}
\subsection{Performance}
\begin{figure}[htb]
    \centering
    \includegraphics[width=0.35\textwidth]{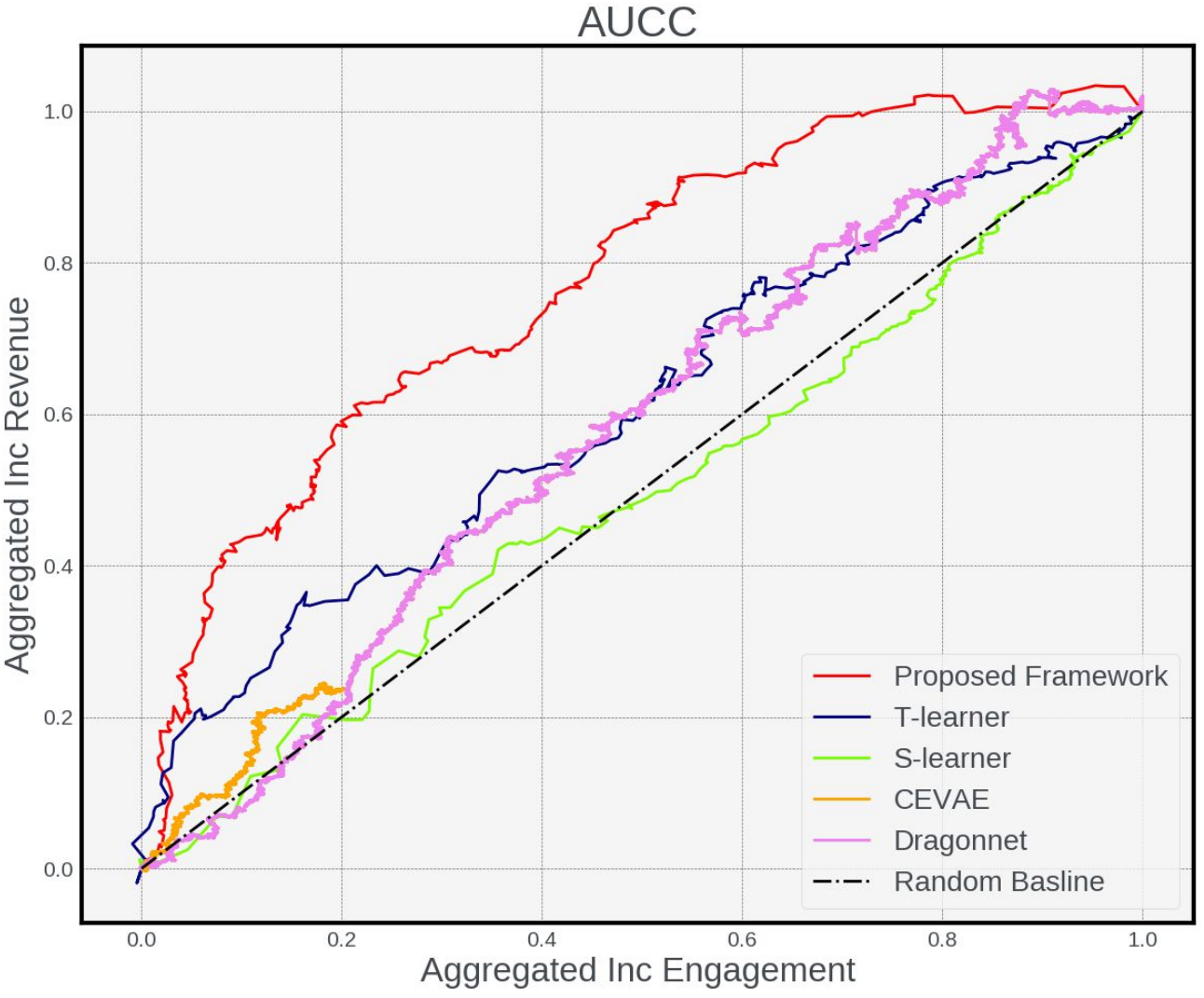}
    \caption{Cost Curves on Product Data.}
    \label{fig:aucc_plot}
\end{figure}

We now demonstrate the effectiveness of our proposed methods by benchmarking all the methods on CRITEO-UPLIFT and the product data. We simulate the product scenario, where outcome predictions are not satisfactory, by training a biased outcome model in CRITEO-UPLIFT benchmark. In the product data, we construct separate CATE models for both revenue and engagement metrics and obtain the resulting a ROI metric of AUCC. 
\\
\textbf{CRITEO-UPLIFT}
We intentionally modify the labels to inject bias into training data for the outcome model. Given the original data $D = \{(X, Y)\}$,
we first deterministically construct a subset according to a feature value filter, $ D' = \{ (X, Y) \mid X \in D, f_0(X) < \alpha \} $, where $f_0$ is a linear function scalarizing the feature vector $X$. We randomly switch the binary label, "visit", from 1 to 0 with a probability of $\beta (\%)$ in the subset. Then we train the outcome model on the entire training dataset with the modified labels. At last, the models trained on the biased training data are evaluated using the original test data.
\\
Table \ref{tab:auuc-criteo} presents the test AUUCs with bias injection at $\beta=0\%$ and $\beta=100\%$. Here, $\beta=0\%$ corresponds to the original training dataset, while $\beta=100\%$ represents the biased training dataset. All methods have shown a consistent high AUUC at $\beta=0\%$, indicating a well-performing outcome model on the original training data. In the case where outcomes are poorly estimated ($\beta=100\%$), our framework still maintains the high AUUC while baselines all suffer from a significant performance degradation.
\begin{table}[h]
\caption{AUUC Comparison on CRITEO-UPLIFT}
\centering
\begin{tabular}{lrr}
\toprule
Methods & $\beta=0\%$ & $\beta=100\%$  \\
\midrule
S-learner & 0.86 & 0.81  \\
T-learner & 0.85 & 0.57  \\
X-learner & 0.87 & 0.72  \\
CEVAE & 0.82 & 0.63  \\
Dragonnet & 0.85 & 0.65  \\
Proposed Framework & 0.86 & \textbf{0.86}  \\
\bottomrule
\end{tabular}
\label{tab:auuc-criteo}
\end{table}
\\
\textbf{Product data}
We evaluate all the methods on the product data in AUCC metrics, as shown in Table \ref{tab:aucc-product}. As the outcomes, long-term revenue and engagement metrics, are difficult to estimate well on their own, all baseline models show unsatisfactory AUCC performances. The proposed framework achieves 20\% improvement over the second-best baseline despite the inevitable misspecified outcome models. The performance improvement stems from the use of propensity score information in a RCT data. Note that DR estimator in \eqref{eq: dr_estimator} is unbiased but has a larger variance than the direct estimator in \eqref{eq: naive_estimator}. The impact of training label variances diminish as dataset size increases. As such, the improvement becomes more pronounced in large-scale applications. Figure \ref{fig:aucc_plot} shows the cost uplift curves of all methods. Here the value and cost in the uplift curves are normalized revenue and engagement, correspondingly. The curve of our framework shows around 40\% revenue gain improvements at most operating points. Fundamentally, our framework leverages the experimental cost of a large-scale RCT and turns it into accuracy improvement in CATE estimates and savings in model complexity.
\begin{table}
\caption{AUCC Comparison on Product Data}
\centering
\begin{tabular}{lrr}
\toprule
Methods & AUCC \\
\midrule
S-learner & 0.51 \\
T-learner & 0.62 \\
X-learner & 0.63 \\
CEVAE & 0.50 \\
Dragonnet & 0.62 \\
Proposed Framework & \textbf{0.77} \\
\bottomrule
\end{tabular}
\label{tab:aucc-product}
\end{table}
Although Dragonnet's training with the regularization yields the estimator satisfying the non-parametric estimating equation, it doesn't show doubly robustness in our product dataset. We also observed the instability during Dragonnet's training on product data. An end-to-end training procedure to achieve doubly robustness in real-world datasets could be left for the future work.

\subsection{Component Analysis}
\begin{figure}[htb]
    \centering
    \includegraphics[width=0.3\textwidth]{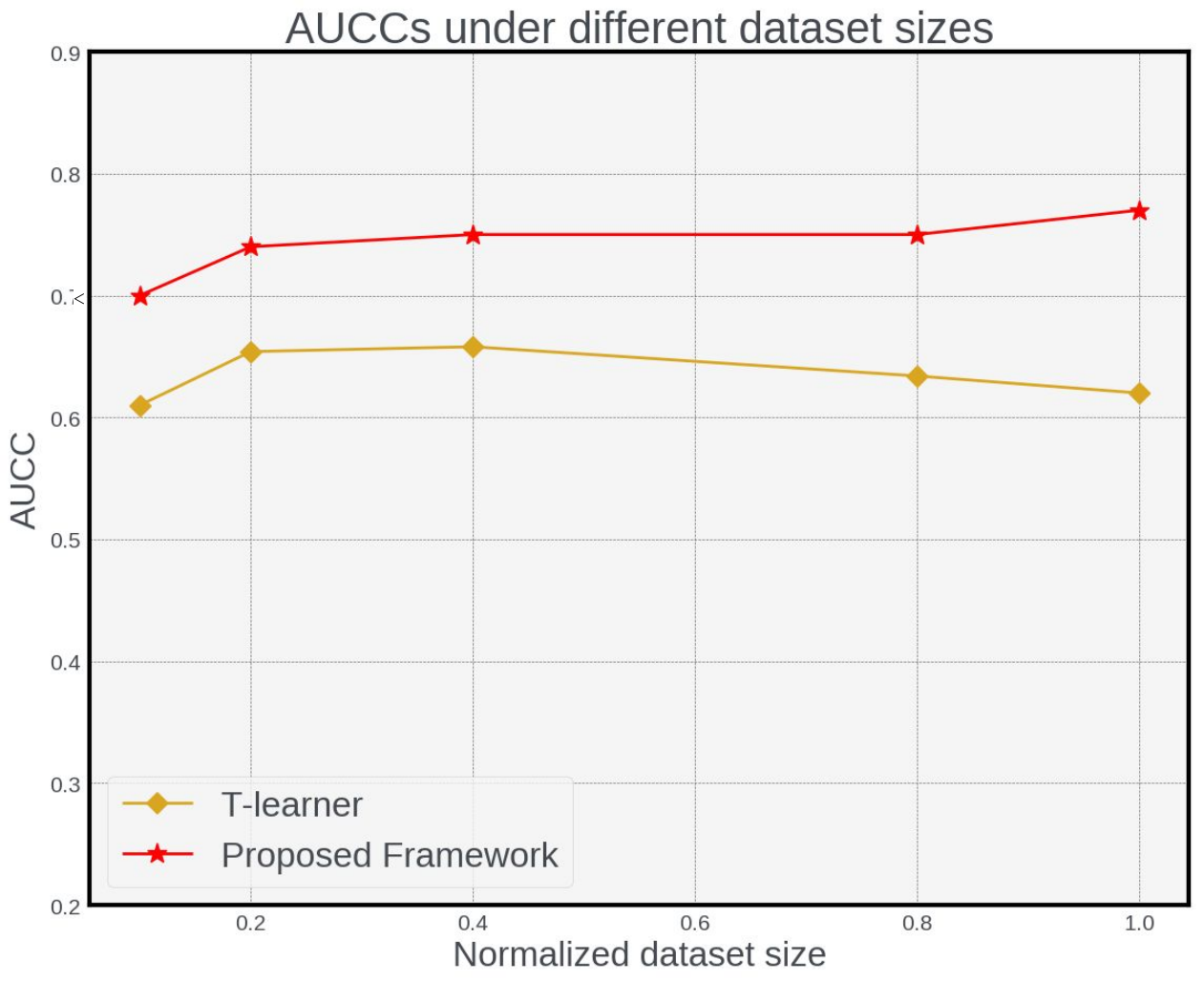}
    \caption{AUCCs with increasing dataset sizes.}
    \label{fig:data_size}
\end{figure}

In this section, we conduct comprehensive experiments to investigate the influence of the propensity score and outcome model on end performance. The resulting insights inform the requirements for our data collection policy and the complexity of our outcome model. Finally, we examine the asymptotic performance of our framework. In some component analysis, the performance of the T-learner has been demonstrated for comparison purposes.
\\
\textbf{Impact of outcome model bias} We adjust the probability of label modification, $\beta$ for CRITEO-UPLIFT, to see how the magnitude of bias impacts different methods differently. 
\begin{figure}[htb]
    \centering
    \includegraphics[width=0.3\textwidth]{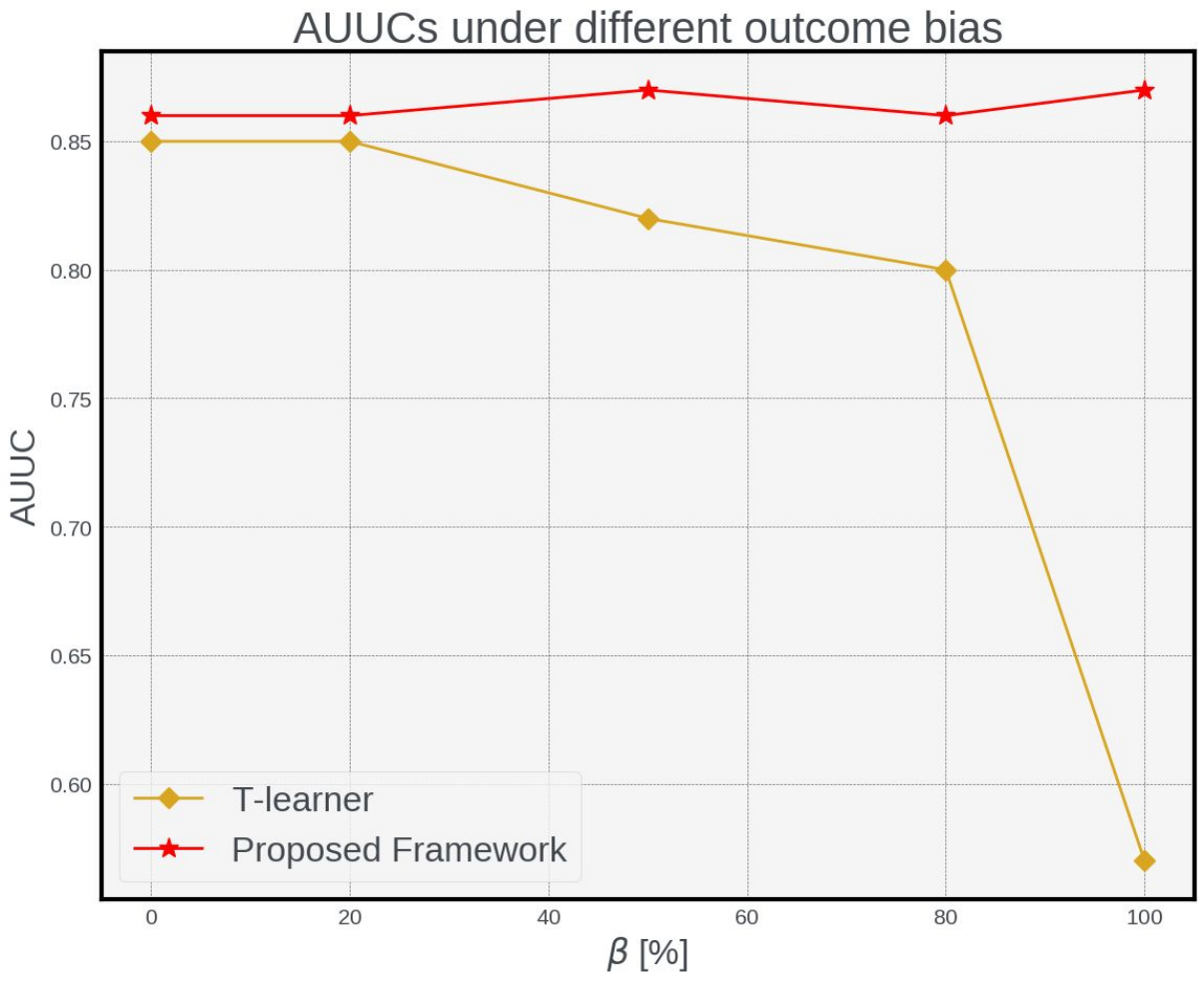}
    \caption{AUUC changes over outcome model bias.}
    \label{fig:outcome_bias}
\end{figure}
As shown in Figure \ref{fig:outcome_bias}, T-learner's AUUC gradually degrades with the increasing outcome model bias. Our framework, on the contrary, maintains a constant AUUC. This AUUC comparison demonstrates that the doubly robust estimator becomes increasingly essential as the bias in the outcome model can directly affect the end result in its absence.
\\
\textbf{Impact of propensity score bias} We adjust the propensity score that is plugged into the doubly robust estimator for CRITEO-UPLIFT. The larger the deviation it has from the ground-truth value, the larger the bias is in the propensity score prediction. The ground-truth propensity score of CRITEO-UPLIFT is 0.85. We offset it by different magnitudes so that we can see how much performance degradation will be caused by the propensity score bias.
\begin{figure}[hb]
    \centering
    \includegraphics[width=0.3\textwidth]{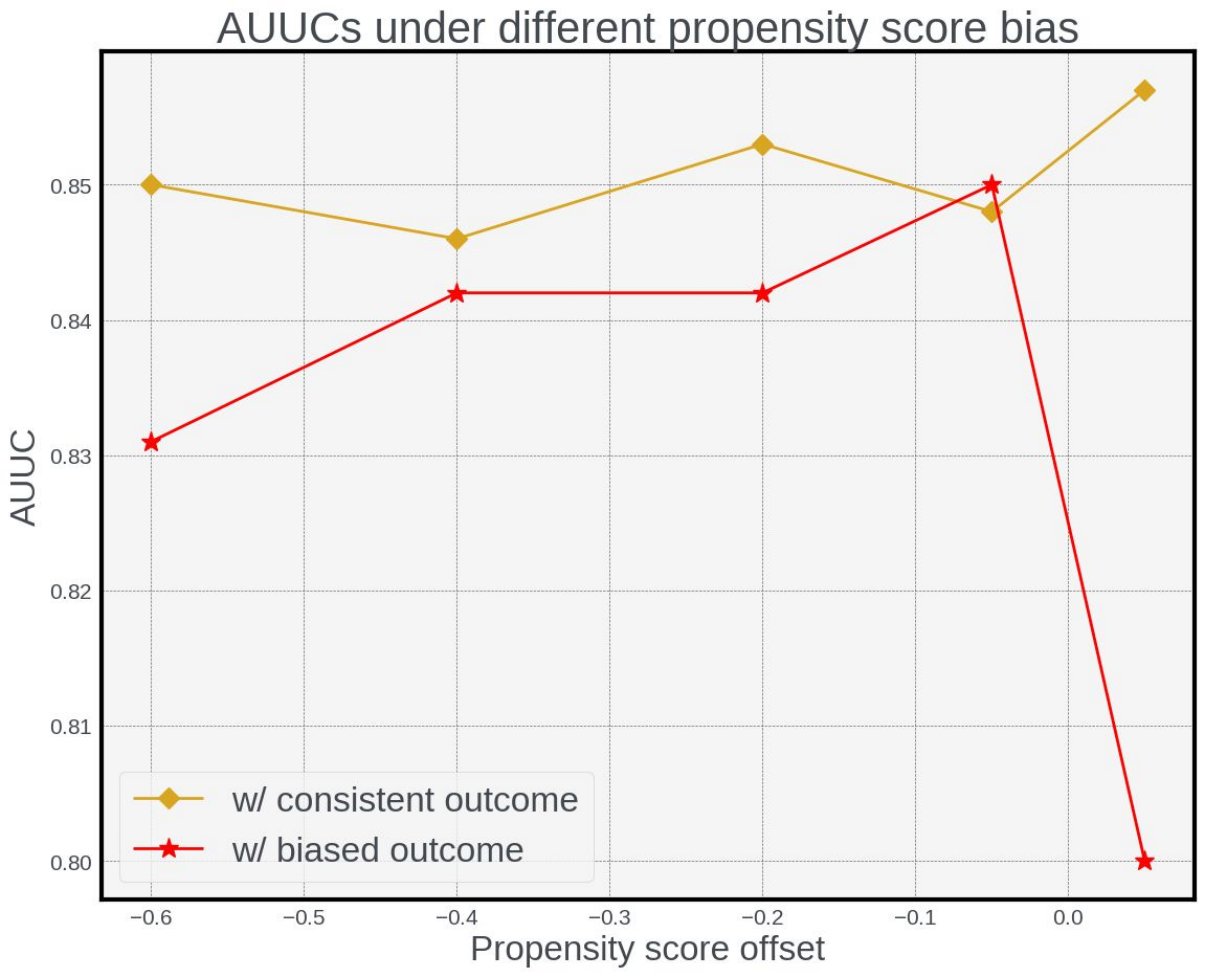}
    \caption{AUUC changes over propensity score bias.}
    \label{fig:prop_score_bias}
\end{figure}
As shown in the Figure \ref{fig:prop_score_bias}, the performance degradation of the proposed method is trivial given the unbiased outcome model. The degradation is larger but still within 5\%. Empirically, propensity score model do not have to be very accurate in our case. Such result suggests that we can further relax our RCT constraint for data collection, especially when the RCT cost is too high.
\\
\textbf{The choice of outcome model} Table \ref{tab:outcome_comparison} shows the AUCCs for the product data with different outcome models. We compare the results of using a constant zero, a random forest, and a MLP model. Improving the outcome model is still beneficial. A better outcome model reduces the variance of doubly robust estimator, leading to a better finite-sample performance in CATE estimates. We can choose an optimal point considering both model performance and complexity.
\begin{table}[h]
\caption{AUCCs with outcome model plugged in}
\centering
\begin{tabular}{lrr}
\toprule
Outcome model & AUCC \\
\midrule
Constant & 0.63 \\
Random Forest & 0.77 \\
MLP & 0.80 \\
\bottomrule
\end{tabular}
\label{tab:outcome_comparison}
\end{table}
\\
\textbf{CATE model performance scaling with data size} Figure \ref{fig:data_size} shows the AUCC trend over the increasing product dataset size. The X-axis is the normalized dataset size. T-learner's AUCC does not show an increasing trend as the dataset size grows larger. On the contrary, our framework's performance is improved as the dataset size scales. Such a comparison indicates that CATE estimates, when solely based on outcome predictions, do not improve with increasing data size. However, the CATE model derived from DR estimator can effectively leverage data scaling. Such scalability because more important as the available training data size grows 
\subsection{Online Result}
We conduct an online A/B test before product launch (one month) and a long-term hold-out test after launch (a couple of months). Long-term revenue gain-to-engagement loss efficiency metrics are used here. The high efficiency indicates that a platform is capable of sustaining profitability over an extended period. The table \ref{tab:online_tests} highlights substantial improvements of +13.8\% and +19.0\% prior to launch, and enhancements of +12.4\% and +20.2\% observed at least 4 months post-launch. In the long term, our deployed work enhances monetization efficiency with live traffic. 
\begin{table}[hb]
\caption{Online Test}
\centering
\begin{tabular}{lrr}
\toprule
Methods & Efficiency Metric 1 & Metric 2  \\
\midrule
Product Baseline & 0\% & 0\% \\
Our work (pre-test) & \textbf{+13.8\%} & \textbf{+19.0\%} \\
Our work (hold-out test) & \textbf{+12.4\%} & \textbf{+20.2\%} \\
\bottomrule
\end{tabular}
\label{tab:online_tests}
\end{table}
\section{Conclusions and Future Works}
In this paper, we develop a doubly robust learner based framework for ads-supply personalization to achieve a sustainable revenue and engagement win. By maximizing the usage of information from the dataset collect through RCT, we significantly improve the user treatment effect estimates, hence a better revenue-to-engagement trade-off. The optimal use of information largely reduces the model complexity. The lightweight framework has been seamlessly integrated with the established ads-delivery system. 
\\
In terms of future directions, devising the method when RCT data are not representative is an interesting challenge. We can gradually extend our methodology from fully controlled data collection to partially controlled and then observational data. In the meantime, we observed the user features drift over time, which requires our future efforts in developing a continuous training for causal models.


\bibliographystyle{ACM-Reference-Format}
\balance
\bibliography{sample-base}

\appendix
\newpage


\end{document}